\documentstyle[12pt]{article}

\def\lp{\left(}
\def\rp{\right)}

\def\trho{\tilde{\rho}}
\def\tp{\tilde{p}}
\def\eku{e^{2(K_{12}-U_{12})}}
\def\eu{e^{2U_{12}}}
\def\-eu{e^{-2U_{12}}}

\def\2om{2\omega +3}

\def\8{8\pi G}

\def\4{4\pi G}
\def\16{16\pi G}

\def\be{\begin{equation}}
\def\ee{\end{equation}}

\begin{document}
\baselineskip.29in

\centerline{\large{\bf Cylindrical thin-shell wormholes and energy conditions}}

\bigskip

\centerline{Claudio Simeone\footnote{Electronic mail: csimeone@df.uba.ar}}

\medskip
\centerline{\it IFIBA--CONICET and Departamento de F\'{\i}sica,}

\centerline{\it  Facultad de Ciencias Exactas y 
Naturales, Universidad de Buenos Aires,}
\centerline{\it Ciudad Universitaria Pab. I, 1428, Buenos Aires, Argentina}

\vskip2cm

\noindent ABSTRACT: We prove the impossibility of cylindrical thin-shell wormholes supported by matter satisfying the energy conditions everywhere, under reasonable assumptions about the asymptotic behaviour of the --in general different-- metrics at each side of the throat. In particular, we reproduce for singular sources previous results corresponding to flat and conical asymptotics, and extend them to a more general asymptotic behaviour. Besides, we establish necessary conditions for the possibility of non exotic cylindrical thin-shell wormholes.

\bigskip





\newpage

\noindent A wormhole configuration connects two regions of spacetime by a throat, thus implying a nontrivial topology, and some consequent interesting features as, for example, the possibility of closed timelike curves \cite{mo,visser,mo-fro}. In the case of wormholes with a compact throat, this is defined as a minimal area surface \cite{visser}, where a flare-out condition is satisfied. The main objection against the actual existence of such wormholes is that, in the framework of General Relativity, they require the presence of exotic matter, i.e. matter violating the energy conditions \cite{visser}.  For wormhole geometries with infinite throats, as cylindrical wormholes are \cite{cilwhs,nos1}, the flare-out condition can be understood in two ways: The areal flare-out condition states that the area per unit length must increase with the radius \cite{nos1}; this leads to the impossibility of fulfilling the energy conditions globally \cite{brle}. However, in Ref. \cite{brle} it was also pointed  that for cylindrical wormholes it may be more appropriate the radial flare-out condition, which only demands that the length of a circunference increases with the radius (see also \cite{nos2,nos3}). Within this approach, wormhole configurations satisfying the energy conditions \cite{brle} were found; however, negative results were obtained in Ref. \cite{brle} for flat and conical asymptotics. Besides, working within the thin-shell class, it was shown that cylindrical wormholes with a positive energy density at the throat are possible \cite{nos3}. However, in Ref. \cite{nos3} we were not able to find solutions completely satisfying the energy conditions; their existence was not discarded, but was left as an open question. Therefore, here we shall address the construction and matter characterization of cylindrical wormholes of the thin-shell class, working under the radial definition of the throat. We shall restrict our analysis to static configurations, and we shall focus on the normal or exotic character of the matter required. We shall prove the impossibility of cylindrical thin-shell wormholes supported by matter satisfying the energy conditions everywhere, under the following assumptions:

\begin{enumerate}

\item  The asymptotic behaviour of the geometries at each side are either flat, local cosmic string-like (i.e. conical) or of the generic Levi--Civita form; they are not necessarily of the same type at both sides.

\item Apart from the required continuity of the line element and a weak condition on the first derivatives of the metric functions (see below), no assumption is made about the metric at the throat.

\item Both metrics have continuous first derivatives outside the wormhole throat (the only shell is at the throat).

\end{enumerate} 
 These conditions include a wide class of generally asymmetric wormholes; symmetric ones, which correspond to equal metrics everywhere at both sides of the throat, are of course included in our analysis as a particular case. The results for flat and cosmic string asymptotics are expected to reproduce those found without recalling singular sources in \cite{brle}. As a corollary, we shall obtain necessary conditions that the throat and asymptotic behaviour of the metrics must satisfy to allow for the existence of non exotic configurations. 

Let us start from two manifolds ${\cal M}_1$ and ${\cal M}_2$ with metrics of the most general  static cylindrically symmetric form \cite{thorne}
\be
ds^2_{12}=\eku(dt_{12}^2-dr_{12}^2) -\-eu W_{12}^2 d\varphi_{12}^2-\eu dz_{12}^2 ,\label{metric}
\ee
where $K_{12}$, $U_{12}$ and $W_{12}$ are functions of the radial coordinates $r_{12}$. For such metrics the Einstein equations relating the geometry with the energy-momentum tensor $\tilde{T}_\mu^\nu=8\pi e^{2(K-U)}T_\mu^\nu={\rm diag}(-\trho,\tp_r,\tp_\varphi,\tp_z)$ take the form
\begin{eqnarray}
\trho_{12} & = &
-{W''_{12}\over W_{12}}+K'_{12}{W'_{12}\over W_{12}} -{U'_{12}}^2,\\
\tp_{r12} & = & K'_{12}{W'_{12}\over W_{12}}-{U'_{12}}^2,\\
\tp_{\varphi 12} & = & K''_{12}+{U'_{12}}^2,\\
\tp_{z12} & = & {W''_{12}\over W_{12}}-2U'_{12}{W'_{12}\over W_{12}}+K''_{12}-2U''_{12}+{U'_{12}}^2,
\end{eqnarray}
where primes denote radial derivatives. From these equations we obtain  the quantities
\begin{eqnarray}
\trho_{12} & = &
-{W''_{12}\over W_{12}}+K'_{12}{W'_{12}\over W_{12}} -{U'_{12}}^2,\label{bulk0}\\
\trho_{12}+\tp_{r12} & = &-{W''_{12}\over W_{12}}+ 2K'_{12}{W'_{12}\over W_{12}}-2{U'_{12}}^2,\\
\trho_{12}+\tp_{\varphi 12} & = & -{W''_{12}\over W_{12}}+K'_{12}{W'_{12}\over W_{12}}+K''_{12},\\
\trho_{12}+\tp_{z12} & = & K'_{12}{W'_{12}\over W_{12}}-2U'_{12}{W'_{12}\over W_{12}}+K''_{12}-2U''_{12},\label{bulk1}
\end{eqnarray}
which should all be positive or at least zero if no exotic matter exists. Now from the two manifolds ${\cal M}_1$ and ${\cal M}_2$ described by (\ref{metric}) we take the regions $r_{12}\geq a$ and paste them at the surface $\Sigma$ given by $r_{12}=a$ to construct a new manifold ${\cal M}$. We assume that the metric functions and the coordinate choice guarantee the required continuity of the line element across $\Sigma$; then the induced metric on $\Sigma $ is unique. We also assume that the flare-out condition is satisfied at $r_{12}=a$, so that the geometry at each side opens up there. Then the manifold ${\cal M}$ constitutes a  wormhole with a matter shell at $\Sigma$, that is a thin-shell wormhole.

 The geometry at both sides of this  surface and the matter on it are related by the Lanczos equations \cite{lanczos}
\be
-[K_i^j]+[K]\delta_i^j=
8\pi S_i^j,
\label{e10}
\ee
where $K_i^j$ is the extrinsic curvature tensor defined by
\be
  K_{ij} = - n_{\gamma} \left. \left( \frac{\partial^2
  X^{\gamma}}{\partial \xi^i \partial \xi^j} + \Gamma_{\alpha \beta}^{\gamma}
  \frac{\partial X^{\alpha}}{\partial \xi^i} \frac{\partial
  X^{\beta}}{\partial \xi^j} \right) \right|_{\Sigma}, \label{e6}
\ee
with $n_{\gamma}$  the unit normal ($n^{\gamma} n_{\gamma} = 1$) to
$\Sigma$ in $\mathcal{M}$. The coordinates of the 4-dimensional manifolds are labeled as $X^\mu=(t,r,\varphi,z)$, while the coordinates on the surface are $\xi^i=(\tau,\varphi,z)$; as usual, $\tau$ stands for the proper time measured by an observer at rest on $\Sigma$.  
The bracket  $[K_i^j]$ denotes the jump ${K_i^j}_2 - {K_i^j}_1$ across the surface $\Sigma$, 
$[K]=\delta^i_j[K_i^j]$ is the 
trace of $[K_i^j]$ and
$S_i^j = {\rm diag} ( -\sigma, 
p_{\varphi}, p_{z} )$ is the surface stress-energy tensor, 
with $\sigma$ the surface energy density and $p_\varphi$, $p_z$  the  surface pressures. For the metrics (\ref{metric}) these equations give the following expressions for the energy density and pressures on the shell:
\begin{eqnarray}
\sigma & = & -{e^{U_2-K_2}\over 8\pi}{W'_2\over W_2}-{e^{U_1-K_1}\over 8\pi}{W'_1\over W_1},\\
p_\varphi & = & {e^{U_2-K_2}\over 8\pi }K'_2 +{e^{U_1-K_1}\over 8\pi }K'_1,\\
p_z & = & {e^{U_2-K_2}\over 8\pi}\lp K'_2+{W'_2\over W_2}-2U'_2\rp+{e^{U_1-K_1}\over 8\pi}\lp K'_1+{W'_1\over W_1}-2U'_1\rp.
\end{eqnarray}
From this we obtain the following relations 
\begin{eqnarray}
\sigma & = & -{e^{U_2-K_2}\over 8\pi}{W'_2\over W_2}-{e^{U_1-K_1}\over 8\pi}{W'_1\over W_1},\\
\sigma+p_\varphi & = & {e^{U_2-K_2}\over 8\pi }\lp K'_2-{W'_2\over W_2}\rp+{e^{U_1-K_1}\over 8\pi }\lp K'_1-{W'_1\over W_1}\rp,\\
\sigma+p_z & = & {e^{U_2-K_2}\over 8\pi}\lp K'_2-2U'_2\rp +{e^{U_1-K_1}\over 8\pi}\lp K'_1-2U'_1\rp.
\end{eqnarray}
In fact, the continuity of the metric across the throat would slightly simplify these expressions, but this is not essential in the subsequent analysis. 

As pointed above, the existence of a wormhole requires that the geometry opens up  (``flare-out condition'') at the throat, that is at $r=a$. The areal flare-out condition would impose
\be
W_2W'_2>0\ \ \ \ \ \ \ \ W_1W'_1>0
\ee
at $r=a$. This requirement automatically implies that the energy density is negative at the shell,  and the energy conditions would be violated (see Ref. \cite{brle} for the same result without recalling singular sources). But the radial flare-out condition only implies
\be
W'_2W_2e^{-2U_2}-U_2'W^2_2e^{-2U_2}>0\ \ \ \ \ \ \ \ W'_1W_1e^{-2U_1}-U'_1W^2_1e^{-2U_1}>0,
\ee
which is less restrictive.  These conditions give
\be
U'_1<{W'_1\over W_1}\ \ {\rm and}\ \ U'_2<{W'_2\over W_2}.\label{fo}
\ee
The condition $\sigma>0 $ ($\sigma=0$ fulfils the energy conditions, but there would be no shell) requires that at least
\be
{W'_1\over W_1}<0 \ \ {\rm or}\ \  {W'_2\over W_2}< 0.\label{sig}
\ee
On the other hand,  the conditions $\sigma+p_i\geq 0$ on the shell impose that at least
\be
 K'_1\geq {W'_1\over W_1}\ \ {\rm or}\ \  K'_2\geq {W'_2\over W_2}\label{p1}
\ee
and simultaneously
\be K'_1\geq 2U'_1 \ \ {\rm or}\ \ K'_2\geq 2U'_2.\label{p2}
\ee
From these results, for a class of metrics with reasonably desirable asymptotics, we shall prove the impossibility of fulfilling the energy conditions everywhere outside the shell if $W'_1/W_1<0$ and $K'_1\geq W'_1/W_1$, or if $W'_2/W_2<0$ and $K'_2\geq W'_2/W_2$. This includes the case in which at both sides at the throat we have $W'/W< 0$, and also the case such that  $K'-W'/W\geq 0$ at both sides. Of course, symmetric wormholes are included in this class too, as long as wormholes which are symmetric at the throat.

\begin{enumerate}

\item Asymptotically flat geometries: in the case of an asymptotically flat behaviour, we must have that when the radial coodinate goes to infinity  it is $e^{2U}\sim 1$, $e^{-2U}\sim 1$, $e^{2K}\sim 1$ and $W^2\sim r^2$. Then very far from the shell we should have $W\sim r$, which implies $W'\sim 1$.

\item Asymptotically cosmic string behaviour: this is the far behaviour associated to a gauge or local cosmic string, and corresponds to a locally flat geometry with a deficit angle. This is given by  $e^{2U}\sim 1$, $e^{-2U}\sim 1$, $e^{2K}\sim1$ and $W^2\sim \alpha^2 r^2$ with $\alpha^2<1$ ($\alpha^2>1$ would describe what is called a surplus angle).   As in the preceding case, at infinity we have $W'\sim 1$.

\item Asymptotically Levi--Civita metrics: in this quite interesting case we have that when we are very far from the shell the metric behaves as $e^{2U}\sim r^{2d}$, $e^{-2U}\sim r^{-2d}$, $e^{2(K-U)}\sim r^{2d(d-1)}$ and $W^2\sim r^2$. Very far from the shell we have $K'-2U'\sim d(d-2)/r$ and $K''-2U''\sim -d(d-2)/r^2$. The Levi--Civita metric becomes invariant under boosts along the $z$ axis for $d=0$ (which gives the flat or conical cases considered above) and $d=2$. The length of a centered circunference increases with the radius only for $d<1$; for $d<0$ the length per unit of $z$ coordinate decreases with  $r$ and vanishes for $r\to\infty$. Thus a reasonable assumption would be $0<d<1$. Under this condition, the far behaviour corresponds to $K'-2U'<0$ and $K''-2U''> 0$. This will be of interest in the analysis presented in the Appendix; however, the hypothesis $0<d<1$ is not central, because in any case the asymptotic behaviour of $W$ is such that $W'>0$, and this constitutes the crucial point in our proof. 

\end{enumerate}
Suppose that at one side of the throat it is $W'/W< 0$ at $r=a$; then according to (\ref{fo}) to fulfill the radial flare-out condition we have $U'<0$ and $|U'|>|W'/W|$. Then if $K'-W'/W\geq 0$ we have $K'W'/W\leq |W'/W|^2<|U'|^2={U'}^2$; hence it must be
\be
K'{W'\over W}-{U'}^2<0.
\ee
 Now, outside the shell the metrics and their derivatives are continuous. This means that the last inequality must be fulfilled by $W$ and $K',\  U',\  W' $ at $r=a^+$, that is in a region immediately beyond the wormhole throat, in the bulk where the energy and pressures are given by Eqs. (\ref{bulk0})-(\ref{bulk1}). So let us assume that the energy conditions are fulfilled at $r=a^+$. From (\ref{bulk0}) this can be possible only if  together with $W'/W<0$ the relation
\be -{W''\over W}>0\label{e}
\ee
 is satisfied at $r=a^+$. But for all the metrics considered, we have the asymptotic behaviour $W'/W>0$. Therefore at some radius $a<r^*<\infty$ we necessarily must have $W'=0$ together with $W''>0$. Then, as $U'^2\geq 0$, the energy density is $\trho_{12}=-W''/W-U'^2<0$ at $r=r^*$ and the energy conditions are violated there.  Note that this analysis does not require equal metrics at both sides of the throat, nor even that the metrics are of the same kind at infinity. One could have, for example, a cosmic string far behaviour at one side, and a Levi--Civita far behaviour at the other side; the steps followed to show that when at the throat there is a shell of normal matter then the energy conditions must be violated beyond it, still apply for different far behaviours. Though cylindrical wormholes supported by normal matter are possible \cite{brle}, any static cylindrical thin-shell wormhole geometry with the throat and asymptotic behaviours considered here requires exotic matter at some finite value of the radial coordinate. Therefore the negative results stated in Ref. \cite{brle} are reproduced here starting from singular sources, and besides, for such matter distributions, are extended to a more general asymptotic behaviour. 

Now let us briefly discuss the complementary point of view. Because $\sigma > 0$ requires at least $W'_1/W_1<0$ or $W'_2/ W_2< 0$ at the throat, let us assume, without loosing generality, that at $r=a$ we have $W'_1/W_1<0$. Then, if we keep  the restriction that $W'_1$ must be positive at infinity, to avoid exotic matter we should require at least that at the throat radius the metric also satisfies $K'_1\leq W '_1/W_1$. Now, the condition $\sigma+p_\varphi\geq 0$ leads to the requirement $K'_2\geq W '_2/W_2$ at the other side of the throat. But if we also keep the condition that asymptotically $W'_2>0$ the analysis above implies that we must admit $W '_2/W_2>0$ at $r=a$, and then $\sigma >0$ forces there the relation $e^{U_2-K_2}|W'_2/ W_2|\leq e^{U_1-K_1}|W'_1/ W_1|$. Finally, the continuity of the metric (i.e. $e^{U_2}/ |W_2| =  e^{U_1}/ |W_1|$ at $r=a$) simplifies the relation to $e^{-K_2}|W'_2|\leq e^{-K_1}|W'_1|$ at the throat. Of course, we have reached this point from the asymptotic behaviour $W'>0$, which is common to flat, cosmic string-like and Levi--Civita metrics.
However, we could relax this condition, and this could be done keeping the desirable property of an ever increasing circunference length, by demanding that $U'<W'/W<0$ asymptotically. Thus, an alternative starting point in the search of non exotic configurations is to assume the latter form for the asymptotic behaviour of the metric. In this sense, our analysis has left us with, at least, necessary relations that should hold between the asymptotic behaviours of the metrics at each side and their first derivatives at the throat so that non exotic matter could support cylindrical thin-shell wormholes. In summary, in relation with the question posed in Ref. \cite{nos3}, here we have a no go result excluding a wide class of metrics as candidates, and we have obtained a guess of the kind of conditions which narrow the search for a positive answer.

\section*{Appendix}

In the case of a Levi--Civita metric with $0<d<1$ we can also carry out the following analysis: If in order to fulfill the energy conditions at the shell we assume that at the same side of the throat it is $W'/W< 0$ and  $K'-2U'> 0$  at $r=a$, then at this radius we have
\be
K'{W'\over W}-2U'{W'\over W}<0.
\ee 
 Besides, the continuity of the first derivatives of the metric implying $K'-2U'> 0$ at $r=a^+$, plus the requirement of $\trho+\tp_z\geq 0$ there, yields 
\be
K''-2U''> 0\label{epz}
\ee
 immediately beyond the throat. But as we have the asymptotic behaviour $K'-2U'<0$, this means that at some radius $r^*>a$ it must be $K'-2U'=0$ together with $K''-2U''<0$, which gives $\trho+\tp_z<0$; that is, at $r^*$ the energy conditions are violated. 

\section*{Acknowledgments}

This work has been supported by Universidad de Buenos Aires and CONICET. I wish to thank E. F. Eiroa for important discussions


\begin{thebibliography}{99}



\bibitem{mo} M. S. Morris and K. S. Thorne, Am. J. Phys. 
\textbf{56}, 395 (1988).

\bibitem{visser} M. Visser, \textit{Lorentzian Wormholes} (AIP Press, New York, 1996). 

\bibitem{mo-fro} M. S. Morris, K. S. Thorne and U. Yurtsever, Phys. Rev. Lett {\bf 61}, 1446 (1989); V. P. Frolov and I. D. Novikov, Phys. Rev. D {\bf 42}, 1057 (1990).

\bibitem{cilwhs} G. Cl\'{e}ment, Phys. Rev. D \textbf{51}, 6803 (1995); G. Cl\'{e}ment, J. Math. Phys. \textbf{38}, 5807 (1997); R. O. Aros and N. Zamorano, Phys. Rev. D \textbf{56}, 6607 (1997); P. K. F. Kuhfittig, Phys. Rev. D {\bf 71}, 104007 (2005).


\bibitem{nos1}  C. Bejarano,  E. F. Eiroa and C. Simeone, Phys. Rev. D \textbf{75}, 027501 (2007); M. Richarte and C. Simeone, Phys. Rev. D \textbf{80}, 104033 (2009).

\bibitem{brle} K. Bronnikov and J. P. S. Lemos, Phys. Rev. D {\bf 79}, 104019 (2009).

\bibitem{nos2}  E. F. Eiroa and C. Simeone, Phys. Rev. D \textbf{70}, 044008 (2004); E. F. Eiroa and C. Simeone, Phys. Rev. D \textbf{82}, 084039 (2010); A. Arazi and C. Simeone, Eur. Phys. J. PLUS {\bf 126}, 11 (2011).

\bibitem{nos3} E. F. Eiroa and C. Simeone, Phys. Rev. D \textbf{81}, 084022 (2010). 

\bibitem{thorne} K. S. Thorne, Phys. Rev. \textbf{138}, 251 (1965).


\bibitem{lanczos} N. Sen, Ann. Phys. (Leipzig) \textbf{73}, 365 (1924); K.
Lanczos, \textit{ibid.} \textbf{74}, 518 (1924); G. Darmois, M\'{e}morial des Sciences Math\'{e}matiques, Fascicule XXV  (Gauthier-Villars, Paris, 1927), Chap. V; W. Israel, Nuovo Cimento B \textbf{44}, 1 (1966); \textbf{48}, 463(E) (1967); P. Musgrave and K. Lake, Class. Quantum Grav. \textbf{13}, 1885 (1996).


\end{thebibliography}
\end{document}